\documentclass[aps,preprint,superscriptaddress,amsmath,showpacs]{revtex4-1}

\usepackage[breaklinks,colorlinks=true]{hyperref}
\usepackage{graphicx}
\usepackage{color}

\begin{document}

\title{
Network of families in a contemporary population: regional and cultural assortativity}

\author{Kunal Bhattacharya}
\email[Corresponding author; ]{kunal.bhattacharya@aalto.fi}
\affiliation {Department of Computer Science, Aalto University School of Science, P.O. Box 15400, FI-00076 AALTO, Finland}
\author{Venla Berg}
\affiliation {Population Research Institute, V\"aest\"oliitto,  Finnish Family Federation, Helsinki, Finland}

\author{Asim Ghosh}
\author{Daniel Monsivais}
\affiliation {Department of Computer Science, Aalto University School of Science, P.O. Box 15400, FI-00076 AALTO, Finland}
\author{Janos Kertesz}
\affiliation {Center for Network Science, Central European University, Budapest, Hungary}
\affiliation {Department of Computer Science, Aalto University School of Science, P.O. Box 15400, FI-00076 AALTO, Finland}
\author{Anna Rotkirch}
\affiliation{Population Research Institute, V\"aest\"oliitto, Finnish Family Federation, Helsinki, Finland}
\author{Kimmo Kaski}
\affiliation {Department of Computer Science, Aalto University School of Science, P.O. Box 15400, FI-00076 AALTO, Finland}
%\begin{figure}
%\centering
%\includegraphics[width=0.75\columnwidth]{Plots/fig-1.eps}
%\caption{\label{fig-1}}
%\end{figure}

\date{\today}

\begin{abstract}
Using a large dataset with individual-level demographic information of 60,000 families in contemporary Finland, we analyse the variation and cultural assortativity in a network of families. Families are considered as vertices and unions between males and females who have a common child and belong to different families are considered as edges in such a network of families. The sampled network is a collection of many disjoint components with the largest connected component being dominated by families rooted in one specific region. We characterize the network in terms of the basic structural properties and then explore the network transitivity and assortativity with regards to regions of origin and linguistic identity. Transitivity is seen to result from linguistic homophily in the network. Overall, our results demonstrate that geographic proximity and language strongly influence the structuring of network. 
%Overall, we find that the patterns of connectivity in the network are heavily influenced by the regions in which the families are rooted. 
%Transitivity is seen to result from linguistic assortativity in family networks. Overall, our results demonstrate the roles of geographic proximity and language in the structuring of family networks. 
%Our results document for the first time the dominance of one single family network in a representative population sample, and demonstrate the roles of geographic proximity and language in the structuring of family networks.

\end{abstract}


\maketitle
\section{Introduction}

Human families include parents, children, grandchildren and lasting pair bonds between usually unrelated spouses, so that families typically encompass at least three family generations and two kin lineages \cite{hughes1988evolution}. This complexity of familial ties allows for various kinds of %complex 
associations between different extended families, for example, through marriage and intermarriage within a kin group. Family members usually help each other by providing emotional, practical and financial support \cite{szydlik2016sharing}. In addition, we learned in a recent study that parents, children, grandchildren and siblings are also known to stay geographically close to each other even in contemporary wealthy and globalised societies \cite{kolk2014multigenerational}, which can lead to genetic homogeneity in certain region. %his spatial proximity may lead to genetic homogeneity in certain regions. 
At the same time, the fertility differences and migration patterns can affect which families contribute most to the overall population. However, there are few studies in which %have investigated 
the structural properties of network of extended families have been investigated.  

Here we investigate the properties of a network of families, using a unique and nationally representative register dataset from contemporary Finland. %, we study the properties of a network of families. 
We investigate the overall network characteristics of the connected components as well as the roles played by the following factors: 
(i) spatial proximity, as measured at %on 
a regional level
(ii) language preferences, as indicated by language (N.B. Finland has two national languages: Finnish and Swedish).
(iii) genetic relatedness, as measured through assumed biological relatedness between horizontal layers of the network.

We have the following research questions:
\begin{enumerate}
\item How is the network of families structured? We identify which regions most of the observed network stems from, as well as the connected components and their geographical origins. 
\item  What does ``clustering'' in the network imply? We explore transitivity within the largest connected components by type of language spoken.
\item How is the network of kins structured? We investigate the influence of the structure of the network of families on the patterns of biological relatedness resulting between individuals within the same generation.
\end{enumerate}

\begin{figure}[h]

\begin{minipage}[b]{0.60\linewidth}
\centering
\includegraphics[width=0.9\linewidth]{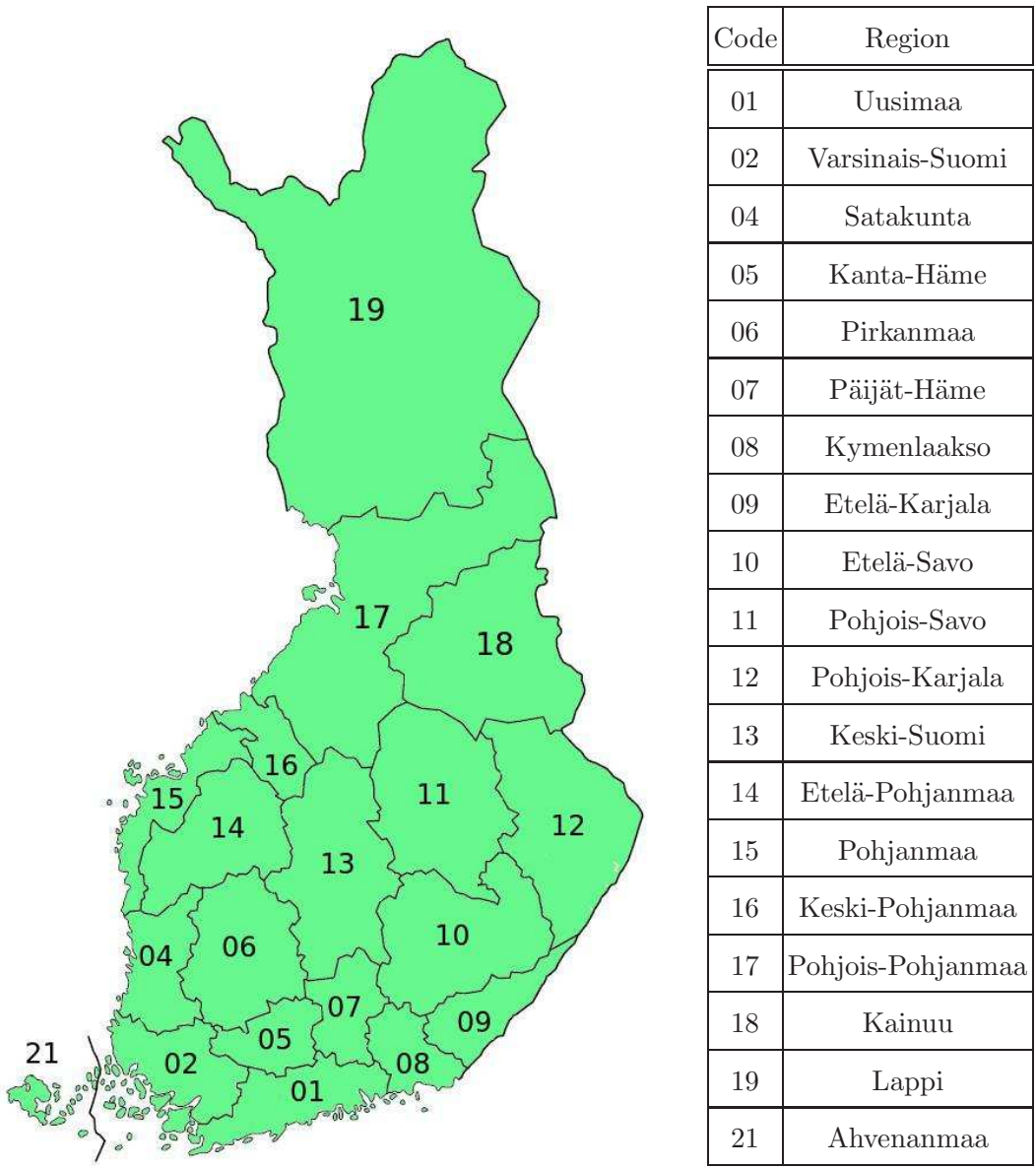}
\par\vspace{0pt}
\end{minipage}%
\begin{minipage}[b]{0.30\linewidth}
\centering
{\small
\begin{tabular}{|c|c|}
\hline
Code & Region\\ 
\hline
\hline
01 &  Uusimaa\\
\hline
02 &  Varsinais-Suomi\\
\hline
04 &  Satakunta\\ 
\hline
05 & Kanta-H\"ame\\
\hline
06 & Pirkanmaa\\
\hline
07 & P\"aij\"at-H\"ame\\
\hline
08 & Kymenlaakso \\
\hline
09 & Etel\"a-Karjala\\
\hline
10 & Etel\"a-Savo\\
\hline
11 & Pohjois-Savo\\
\hline
12 & Pohjois-Karjala\\
\hline
13 & Keski-Suomi\\
\hline
14 & Etel\"a-Pohjanmaa\\
\hline
15 & Pohjanmaa\\
\hline
16 & Keski-Pohjanmaa\\
\hline
17 & Pohjois-Pohjanmaa\\
\hline
18 & Kainuu\\
\hline
19 & Lappi\\
\hline
21 &  Ahvenanmaa\\
\hline
\end{tabular}
}
\par\vspace{0pt}
\end{minipage}
\caption{Map showing the $19$ different administrative regions (``Maakunta'') of Finland \cite{tilastoskesus2015handbook}. The names are provided on the right with the corresponding codes \cite{statfi}.}
\label{map}
\end{figure}

\section{Data}
The dataset is a nationally representative and anonymised dataset of multiple generations of individuals of the late 20th century population of Finland (with the recent census of about 5.5 million), derived from the National Population Register of Finland through Statistics Finland. The data consists of 60,000 randomly selected Finns (index-persons) from six birth cohorts (1955, 1960, 1965, 1970, 1975, and 1980),  each having about 10,000 people constituting 11 -- 16 per cent of the total cohort. This dataset, consisting altogether 677,409 individuals, including the index persons' parents and parents' other children, i.e., siblings and half-siblings as well as the index persons' and their (half-)siblings' children and children's children. In the case of half-siblings, the data includes the half-sibling's other parent, either mother or father (randomly selected), to avoid including two half-siblings that are not genetically related. Thus the data comprise extended families of four generations: the zeroth generation comprising of mothers and fathers; the first generation comprising of index-persons and their siblings and half-siblings; the second generation comprising of the children; and the third generation comprising of the grandchildren. We could thus also separate between cousins and second cousins within the same horizontal family generation. 

For each individual, the data has demographic information including time of birth,  place of birth (administrative regions called ``Maakunta'' in Finnish), time of death, time of marriage (and divorce), yearly information of the place of residence (region). The currently demarcated administrative regions or Maakunta's in Finland exhibit a substantial degree of cultural and economic similarity including recognizable regional dialects, symbols and local food traditions \cite{virrankoski2001suomen}. For our analysis we consider here $18$ out of the $19$ regions, excluding Ahvenanmaa (the {\r A}land Islands) region owing to its small population and being separated from the mainland of Finland. % the last region (the {\r A}land Islands) is left out owing to its small population and being separated from the mainland. 
A few regions stand out historically and culturally: Uusimaa is the region of Finland's capital Helsinki and the largest urban settlement in the country: 30 per cent of the total population lives in Uusimaa. The former capital Turku is now the third largest city and located in Varsinais-Suomi region, while the second largest city Tampere is situated in the region of Pirkanmaa. %Southwest Finland, while the third largest city Tampere is situated in the region of Pirkanmaa. 
The region in the middle and Western Finland, in the various Pohjanmaa regions, %Ostrobotnia, 
is known for a history of agriculture and entrepreneurship, and also for its comparatively high fertility. A religious sect within the Protestant church, called the Laestadians live in this area, especially in the Pohjois-Pohjanmaa region. Laestadians do not approve of modern contraception, which has contributed to a larger proportion of large families with four or more children both among the members of this sect and also among their non-Laestadian neighbours \cite{terama2010regional}. Also many regions and especially the Northern and Eastern regions of Finland have witnessed emigration to the Southern regions, especially to Uusimaa region \cite{2017arXiv170802432G}. Finally, Finland has a national minority of Swedish-speaking Finns, comprising around 6 per cent of the total population. Swedish speaking Finns are typically living in the Western and  
coastal areas or regions of the country.

\section{Methods}
We construct the network using the data of the extended families of the index-persons. A node in the network is a `family' comprising of an index individual and his or her parents, full siblings and half-siblings, children and grandchildren, reflecting the four  family generations in our data as described above. A link between two families is a `parental union', defined as a male and a female who are married and who have at least one common child.  We identify the links by searching for individuals that belong to multiple families. Note, that the presence of a such a person in a given family would also mean the presence of one of the parents (mother or father), with whom the person is related genetically. Thus identifying such persons in turn allowed us to identify links between families -- parental unions consisting of a male and female, each from a different family, whose union has resulted in one or more offspring. Once such a parental union was found we attributed the year of birth of the first offspring to this union. Together, the set of parental unions (links) and the sets of families (nodes), constitute the network of families.

Additionally, we assign to each family, a reference year and a region of origin. The reference year, taken here as the year of birth of the index person, allows for a gross comparison between the generation of individuals belonging to different families. %The reference year is taken as the year of birth of the index person. 
Our aim is to study the parental unions that link different families. Therefore, for a given family we focus on the birth regions of those individuals who have children. (We exclude the generation 0 mothers and fathers, as by definition they belong to the same family). However, not all individuals in a given family will have the same birth region. In cases when the birth regions of the reproducing adults in a single family are extremely diverse, the assumptions with regard to the regional influences become weak. In contrast, %On the contrary, 
the number of families where all the reproducing adults are born in the same region is expected to be smaller. Therefore we calculate the number of families in which at least a fraction $\theta$ of its reproducing adults are born in the same region (see the Appendices). We choose $\theta=0.6$ which allows to include $81\%$ of of the original number of families, while also fulfilling the criterion of having a large majority of the reproducing adults in a given family being born in the same region, assigned as the region of origin for the family. We assume that the region assigned to a family has over time influenced the different generations at multiple levels including social, cultural and genetic inheritance, so that transitivity and assortativity may further intensify the cultural and genetic density. 

The index persons were chosen randomly from the whole population of Finland. In this sense, the network that we construct is sampling of the real network in place. However, the features that emerge from sampling seem to have resulted from the influences of diverse regional factors. In this sense the sampled network provides us with the ``lower limits'' on the structural characteristics present in the actual network. 

\begin{figure}[!ht]
\centering
\includegraphics[scale=0.45]{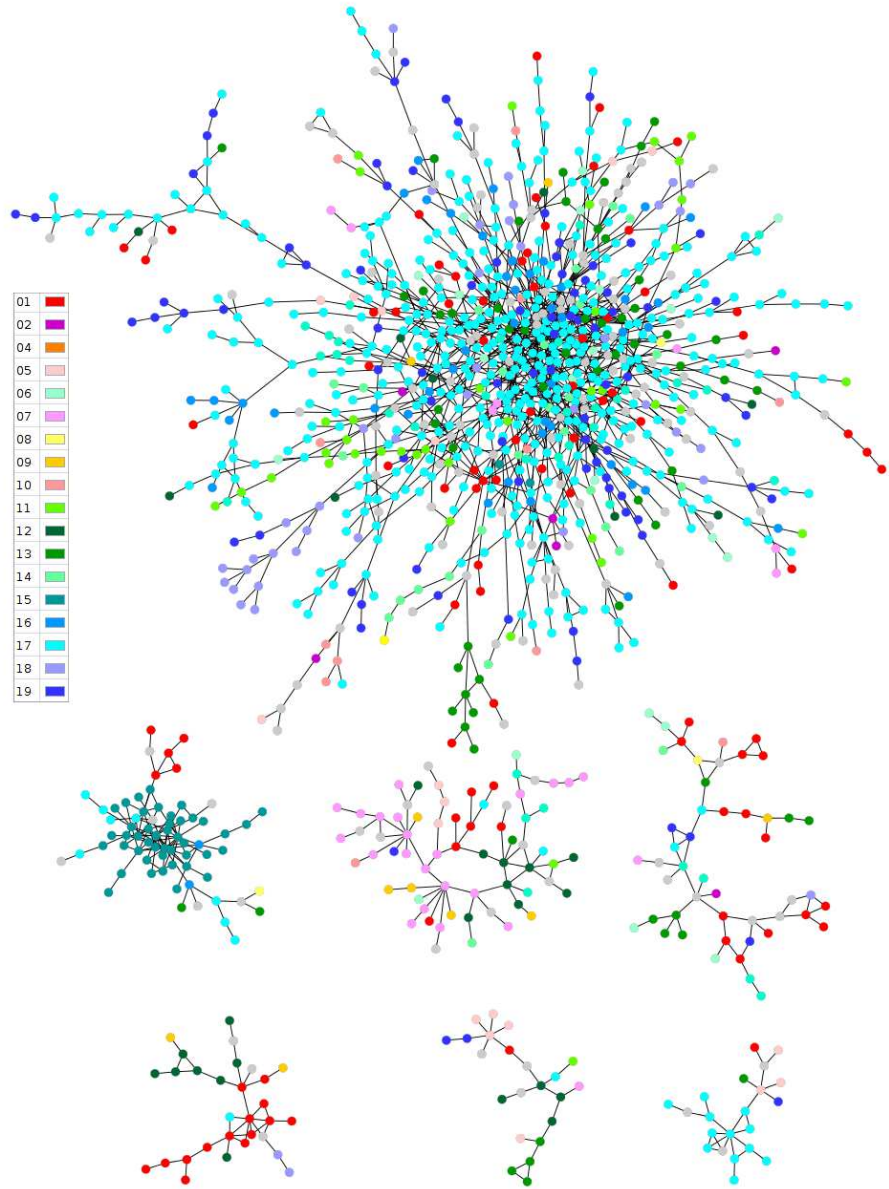}
\caption{{\bf The network of Finnish families.} Nodes represent families, links between two nodes are associations with off-spring common to both families. The first seven components of the network are shown in descending order of sizes. The largest connected component (LCC) has $957$ nodes and $1211$ links. The next in the series have $68$, $68$, $49$, $30$, $22$ and $21$ nodes, respectively. The different colours represent different regions of origin for the families and are indicated in the legend. The families that could not be associated with a particular region are denoted in grey.}
\label{network-paper}
\end{figure}

\section{Results}

\begin{figure}
\centering
\includegraphics[width=0.85\columnwidth]{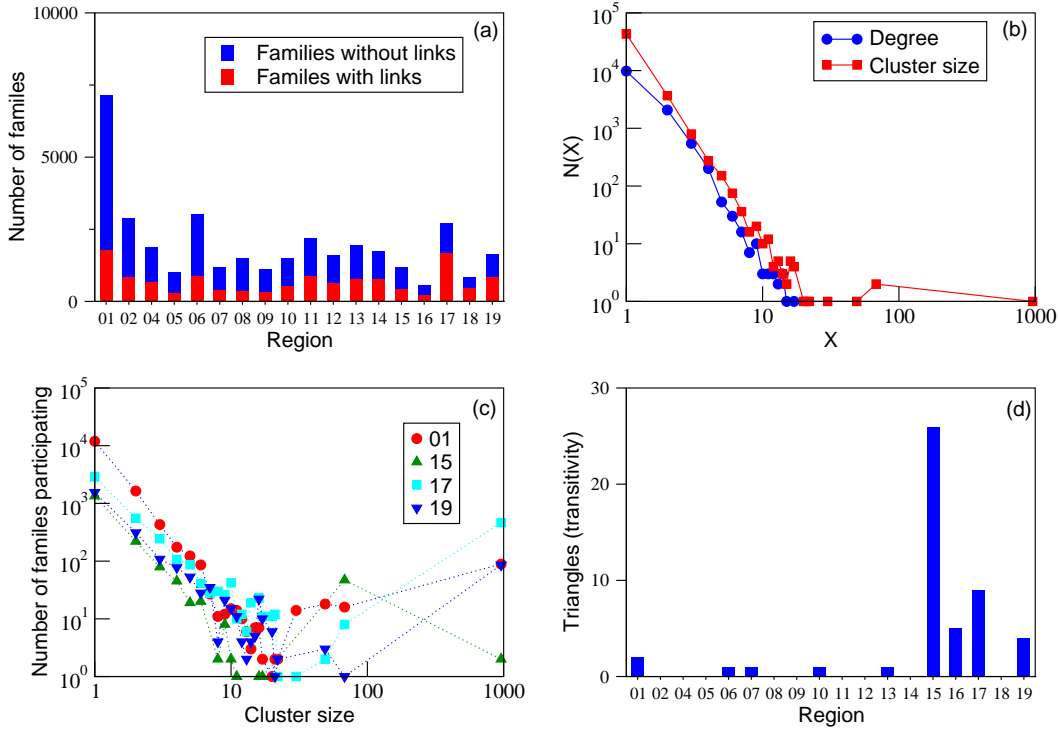}
\caption{{\bf Characteristics of the network.} (a) Number of families distributed over the $18$ regions of mainland Finland.  The plot shows the number of families that could be linked to other families in comparison with the total number of families. (b) The degree distribution of the network (circles) and the distribution of sizes of connected components (squares). (c) Participation in the different components by families from different regions (four cases illustrated). (d) Number of closed triplets of nodes (triangles) in each region. }
\label{network-stat}
\end{figure}

\subsection{Connected components and regions of origin}

First, we discuss our findings related to the structure of the network of families. Of the $60,000$ families in the network, for $12,754$ we could detect linkages to other families in the network. The total number of links (parental unions) is $8,648$. The families with linkages produce a set of connected components with the largest connected component (LCC) being made up of $957$ families and $1,211$ links. Fig~\ref{network-paper} illustrates the first seven components, in descending order of size. We can see that the LCC is much larger than the following components, with 957 nodes in the LCC compared to 68 nodes in the two components next is size. %This indicates that our representative population sample features one single, huge kin network which includes around one tenth of the population (8 percent of the detected nodes and 14 percent of the detected links). 
The distribution of the size  of the connected components and the degrees (number of connections for a given node) are shown in Fig.~\ref{network-stat}(b). The degree distribution decays very fast terminating at a maximum degree of $17$. The distribution of clusters in the network is similar in shape but indicates the presence of the LCC and other larger clusters. 
The composition of the clusters can be analyzed to show how the families from any given region are distributed among them. For a given cluster size we plot the number of families from a given region. Fig.~\ref{network-stat}(c) shows the results for four different regions. Interestingly, this figure reveals that within the LCC (extreme right on the domain) the Pohjois-Pohjanmaa region %Northern Ostrobotnia 
contributes most, followed by an almost equal presence of regions of Uusimaa) and Lapland. 

Thus the LCC is dominated by families from a particular region. Overall, for $13\%$ of the families that have links, the region of origin is Pohjois-Pohjanmaa region. %Northern Ostrobothia. 
This is a large contribution since the largest number for region of origin is $14\%$ of all families, coming from the Uusimaa region around the capital Helsinki. However, the dominance by Uusimaa region is visualised on the extreme left, in the case of the smaller clusters (including families that could not be linked), where the largest contribution is from this region. This comparison is shown in Fig.~\ref{network-stat}(a). Indeed, the total number of families (with or without links) that originate from Uusimaa region make up a much larger fraction of the total set of families when compared to any other region. The dominance of the Uusimaa region is not surprising, since as mentioned in the Data section the Uusimaa region includes the capital area and it has the highest population size and density of all the regions of Finland.

\begin{figure}
\centering
\includegraphics[width=0.75\columnwidth]{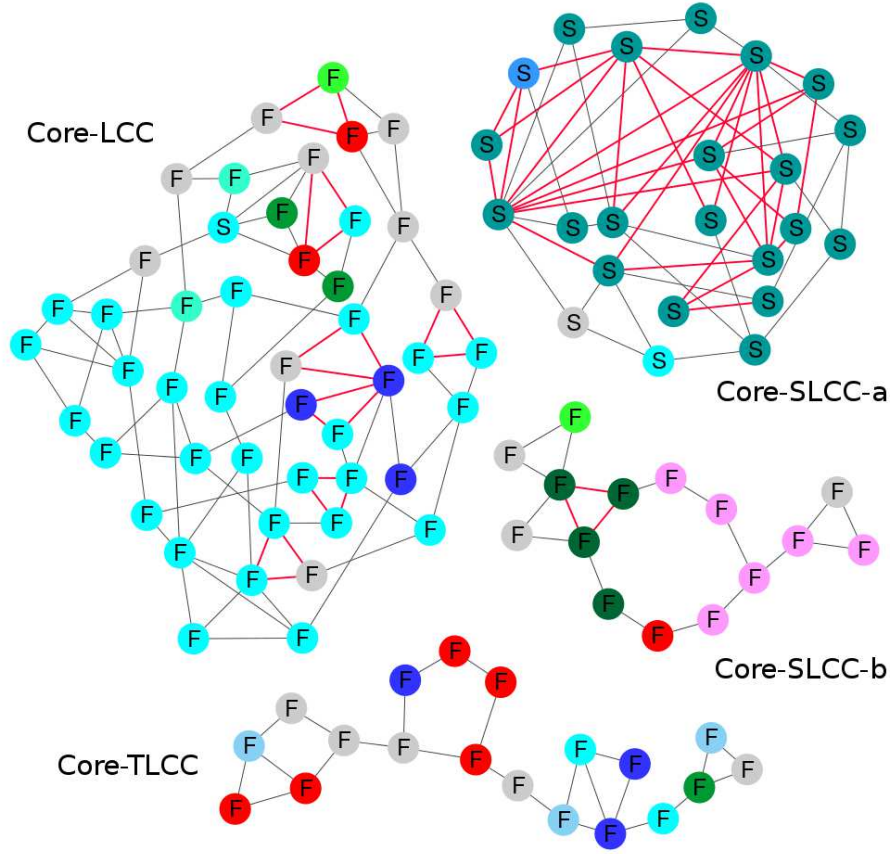}
\caption{{\bf Presence of transitivity.}  Four different dense subgraphs with variation in the proportion of transitive triangles. Links that are part of transitive triangles are coloured in red. The label on a node indicates the language that is spoken by the majority of the adult individuals in the family -- Finnish (`F') or Swedish (`S'). Core-LCC: The $k_\textrm{max}=3$ core of the LCC ($N=47$, $L=86$). This subgraph is dominated by nodes from region 17 (Pohjois-Pohjanmaa). Core-SLCC-a: The $k_\textrm{max}=4$ core extracted from the first of the two second largest connected components as they appear in Fig. \ref{network-paper} ($N=23$, $L=55$). Most of the nodes in this subgraph belong to region 15 (Pohjanmaa). Core-SLCC-b: The $k_\textrm{max}=2$ core extracted from the second of the two second largest clusters ($N=15$, $L=19$). Core-TLCC: The $k_\textrm{max}=2$ core extracted from the third largest cluster ($N=19$, $L=22$).}
\label{triads-lcc}
\end{figure} 

\subsection{Transitivity}

For the second research question we investigated transitivity by considering the triangles that may reflect the transitivity with regard to family relations \cite{wasserman1994social}. In Fig.~\ref{network-stat}(d) we plot the the number of transitive triangles (Fig. \ref{triads}) that families from each region participates in. The highest amount of triangulation is found to occur in the region 15 (Pohjanmaa). %(Ostrobothnia). 
Expecting that the presence of triangulations would lead to an increase in the number of linkages in the neighbourhood of the corresponding nodes, we probe the strongly connected regions of the network. 

We extract the components by performing a $k$-core decomposition \cite{dorogovtsev2006k}. For the LCC we find that, $k_\textrm{max}$, the maximum value of the degree ($k$) for which a core exists is $3$ (i.e. a family belonging to the core is connected to three or more families). Therefore, the full LCC with $957$ nodes could be partitioned into $3$ shells. The outermost shell (a family has atleast one connection) has $600$ nodes, the shell in the middle (a family has atleast two connections) has $310$ nodes, and the central core has $47$ nodes. While as a whole the LCC has an average degree of $2.5$, the value at the core is $3.7$. In addition, %Also, 
the concentration of families belonging to region $17$ (Pohjois-Pohjanmaa) %Northern Ostrobotnia 
increases from the outermost shell ($44\%$) to the core ($58\%$). 

We obtain the cores of the four largest components of the network, as depicted  in Fig.~\ref{triads-lcc}. Whereas, the core of the LCC (Core-LCC) is dominated by families from the region 17 (Pohjois-Pohjanmaa), as described above, the core of one of the two second largest clusters (Core-SLCC-a, $k_\textrm{max}=4$) is found to be composed of families mainly from the region 15 (Pohjanmaa). %(Ostrobothnia). 
The presence of a large number of transitive triads is observed in the latter core with links depicted in colour red. Such triads are also visible in the Core-LCC, but %however, 
their number is low. The other two cores (Core-SLCC-b and Core-TLCC, having $k_\textrm{max}=2$) are relatively smaller in size and do not show the presence of triads.

To understand the possible reasons behind the simultaneous presence of such large number of triads in a specific subgraph (Core-SLCC-a) we probe the cultural similarities between the families. We labelled the nodes based on the language spoken by the majority of the adults in these families. Results show (Fig. \ref{triads-lcc}) that the Core-LCC is dominated by Finnish speakers, featuring only one family with a majority of Swedish speakers whereas, the Core-SLCC-b and the Core-TLCC have none at all. By contrast, the core-SLCC-a has exclusively families with a majority of Swedish speakers. This indicates a very high degree of linguistic attraction within a population among both Finnish and Swedish speakers, and a higher degree of intermarriage and genetic relatedness among the Swedish-speaking cluster (Core-SLCC-a), as reflected in the frequency of transitivity.

\begin{figure}
\centering
\includegraphics[width=0.85\columnwidth]{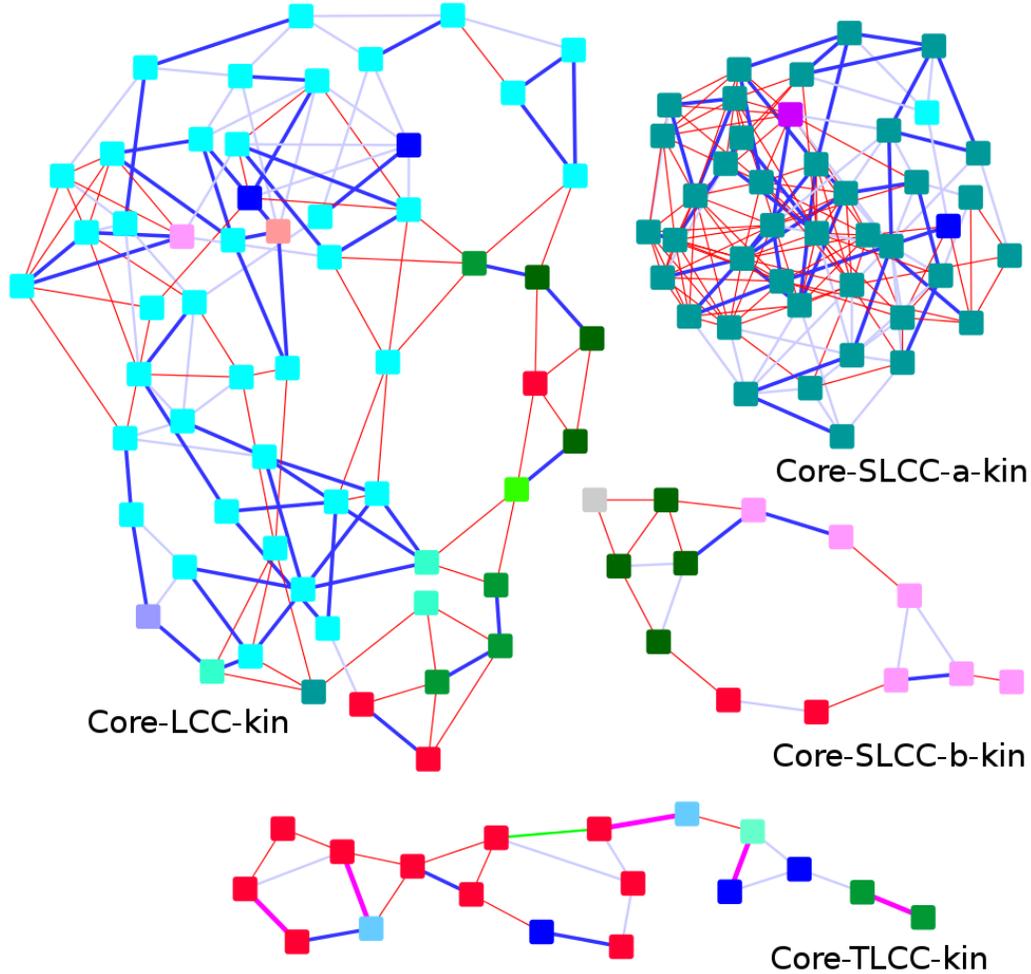}
\caption{{\bf Network of kins.}  The subgraphs represent patterns of connection between individuals who are horizontal kin and are extracted from the network of families illustrated in Fig. \ref{triads-lcc}. Each node in the above subgraphs  is an individual (offspring) resulting from a link (parental union) in the subgraphs in Fig. \ref{triads-lcc}. The colour of a node represents the birth region of the individual. The colour of a link indicates the nature of kinship -- magenta: half siblings, blue: first-cousins, violet: first-cousins-once-removed, green: half-first-cousins, red: second-cousins. We denote the kin network corresponding to the core of the LCC as Core-LCC-kin, that corresponding to the core of first of the second largest connected component as Core-SLCC-a-kin,  the kin network corresponding to the core for the other second largest connected component as Core-SLCC-b-kin, and the core for the third largest connected component as Core-TLCC-kin. The characterization of these subgraphs in terms of the usual topological parameters is provided in Table \ref{tab:kin-stat}.}
\label{kin-graph}
\end{figure} 

\begin{table}
\caption{{\bf Parameters describing the structure of the subgraphs in the network of kins:} number of nodes ($N$), average degree ($\langle k\rangle$), average shortest path length ($d$), clustering coefficient ($CC$), the average relatedness on links ($\langle r\rangle$, in $\%$), and the average aggregated relatedness for nodes ($\langle r_\textrm{sum}\rangle$). The average relatedness is calculated from the fact that each link corresponds to one of the following values ($\%$) of assumed genetic relatedness that are, $25.0$ for half siblings, $12.5$ for first-cousins, $12.5$ for half-aunt/uncle-niece/nephew, $6.25$ for first-cousins-once-removed, $6.25$ for half-first-cousins, $3.125$ for second-cousins, and $1.5625$ for half-second-cousins. We also provide ratios $d/d_\textrm{random}$ and $CC/CC_\textrm{random}$ where, $d_\textrm{random}$ and $CC_\textrm{random}$ are the average shortest path length and clustering coefficient for the Erd\H{o}s--R\'enyi model having same $N$ and $\langle k\rangle$.}

\begin{tabular}{l@{\hskip 0.2in}*{10}{c@{\hskip 0.2in}}c}
\hline
Subgraph              & $N$ & $\langle k\rangle$ & $d$ & $d/d_\textrm{random}$ & $CC$  & $CC/CC_\textrm{random}$ &   $\langle r\rangle$ & $\langle r_\textrm{sum}\rangle$\\

\hline
\hline
Core-LCC-kin    & 58 & 4.5 & 3.6 & 1.3 & 0.43 & 5.4 &   7.2 & 32.5\\
Core-SLCC-a-kin & 42 & 8.6 & 2.0 & 1.2 & 0.53 & 2.5 &  5.7 & 49.5 \\
Core-SLCC-b-kin & 13 & 2.8 & 2.7 & 1.1 & 0.32 & 1.4 &  5.4 & 15.4 \\
Core-TLCC-kin & 18 & 2.6 & 3.8 & 1.2 & 0.21 & 1.4 &  10.0 & 26.0\\
\hline
LCC-kin & 1052 & 5.5 & 6.9 & 1.7 & 0.54 & 108.0 &  7.2 & 40.3\\
SLCC-a-kin & 93 & 8.1 & 3.0 & 1.4 & 0.57 & 6.3 &  6.3 & 51.0 \\
SLCC-b-kin & 67 & 4.1 & 4.9 & 1.7 & 0.46 & 7.7 &  6.8 & 27.9 \\
TLCC-kin & 44 & 3.4 & 5.0 & 1.6 & 0.53 & 6.6 &  8.1 & 27.5 \\
\hline
\hline
\end{tabular}
\label{tab:kin-stat}
\end{table}

\subsection{The network of horizontal kin ties}

For our third research question we study the pattern of genetic relatednesses within the available generations of the extended families by constructing the kin network. Each node in the kin network is an individual (firstborn offspring) resulting from a link (parental union) in the network of families. Therefore, a node in the network of families having  distinct links to two different families results into a pair nodes (individuals) in the kin network that are linked (kinship). For example, in a case when two sisters from a given family marrying into two different families, the two firstborns become linked as first cousins. Assuming that all families are distinct in terms of the genetic material that is inherited by its members, we only include those kinship relations where the sets of families of the two individuals do not completely overlap. Thus, kinships such as the parent-offspring relationship are excluded from the study. The birth cohorts of the individuals in the kin network being restricted by the facts that around $80\%$ of the links in the network of families appear in a span of 20 years (Fig. \ref{link-year}) and $50\%$ within 10 years, allow us to term the network as a network of horizontal kin ties.

First we extract the kin graphs shown in Fig. \ref{kin-graph} from the four cores shown in Fig. \ref{triads-lcc}. The kin graphs corresponding to the cores of the family graphs reveal compositions very similar to the family graphs themselves in terms of the birth regions of the individuals. The dense linking found between the families in the Core-SLCC-a is converted into a clustering between kins in the Core-SLCC-a-kin. The different parameters characterizing the structures of these kin graphs are summarized in Table \ref{tab:kin-stat}. For the average shortest path length ($d$) and the clustering coefficient ($CC$), we provide values corresponding to a Erd\H{o}s--R\'enyi model for random linkages  with similar values for number of nodes and edge-density. We also use the information of the types of kinship to measure the following quantities. We calculate the average coefficient of relationship $\langle r\rangle$ by summing over all the genetic relatednesses for all the links in a given subgraph and then dividing by the total number of links in the subgraph. We also provide $\langle r_\textrm{sum}\rangle=\langle k\rangle\dot\langle r\rangle$, which is the average of aggregated genetic relatedness at nodes.  

Among the four kin graphs corresponding to the cores in the network of families, the $CC$ appears to be the highest in the Core-SLCC-a-kin, and as such results from the transitive triangulations observed in the Core-SLCC-a composed of Swedish speaking families. In the Core-SLCC-kin, in contrast to the rest three, the fact that a random individual could be found linked to the highest number of close kins is evidenced from the high values of the average degree $\langle k\rangle$ and the average aggregated genetic relatedness $\langle r_{sum}\rangle$. Interestingly, the average relatedness in the network appears to be high in Core-TLCC-kin, which is due to the presence of half-sibling relationships. Under the criterion, $d/d_\textrm{random}\gtrsim 1$ and $CC/CC_\textrm{random}\gg 1$ \cite{watts1998collective}, all the four graphs appear to be small worlds in terms of structure. 

Additionally, we include in the analysis the kin graphs directly derived from the four largest clusters in the network of families (without being restricted to their cores). Remarkably, for the LCC-kin, the kin graph corresponding to the LCC in the network of families, which is far more larger in size compared to the LCC-core-kin ($N=1052$ and $N=58$), the small world character appears to be preserved if not enhanced as observed from the amplification of ratio $CC/CC_\textrm{random}$ with only marginal increase in the value of $d/d_\textrm{random}$.

\begin{figure}
\centering
\includegraphics[width=0.95\columnwidth]{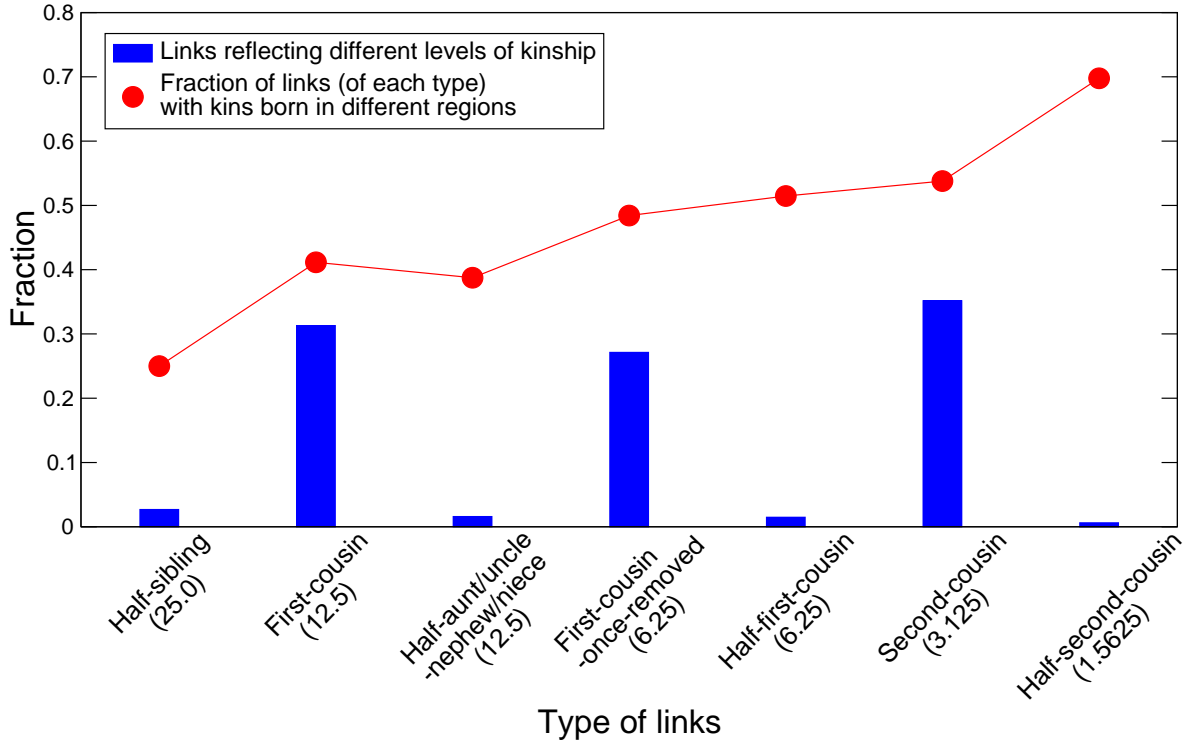}
\caption{{\bf Nature of kin ties.} The fraction of different types of links present in the network of kins is indicated by the bars. For a given type of kinship, the circles indicate the fraction of cases (out of the total in each case) when the individuals in the pair were born in different regions.}
\label{kin-graph-stat}
\end{figure}

In Fig. \ref{kin-graph-stat} we show the frequencies of different types of kinships that are found in the entire kin network network. The relationships that are most abundant (around $30\%$ in each case) are the first--cousin, first--cousin--once-removed and second--cousin. Relationships that mainly originate from family ties formed due to multiple marriages of individuals are present in smaller number. For each kind of relationship we also provide the fraction of cases where the individuals (kins) are born in different regions. This fraction is found to increase as the tie strength (characterized by the genetic relatedness) decreases. A possible cause behind this is the migration of individuals from single extended families into the different regions of over long time scales spanning over generations.

\subsection{Assortativity}

\begin{figure}
\centering
\includegraphics[width=0.95\columnwidth]{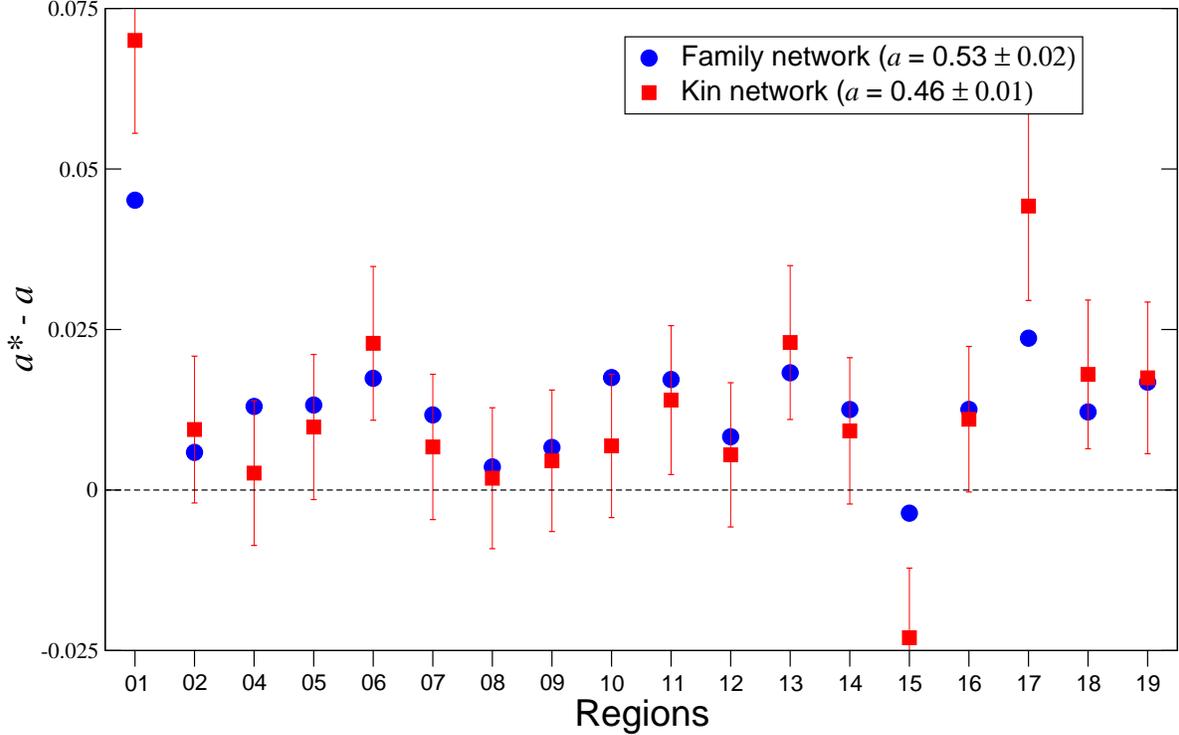}
\caption{{\bf Assortativity coefficient.} The assortativity coefficient ($a$) is calculated for the network of families using the region of origin of families (circles), and for the network of kins using the birth region of individuals (squares) (see Appendices). The values of $a$ are indicated in the legend. Then for a given network, all the nodes belonging to a particular region are removed, and the assortativity coefficient is recalculated ($a^*$). The difference $a^*-a$ with respect to particular region (on the horizontal axis) indicates the effect of having such nodes in the network. For example, a value of $a^*-a$ significantly greater than zero, indicates that these nodes when present in the network show a tendency to get connected nodes from other regions. For the kin network, the error bars in the quantity $a^*$ are shown. For the network of families, we observe a clear correspondence of the values of $a^*-a$ with the values in case of the kin network. However, the errors are much larger in magnitude (due to a lower link density) and therefore, are not shown in the figure.}
\label{assortativity}
\end{figure}

Finally, we characterize the network of families as well as the network of kins in terms of the assortativity coefficient. In general, this coefficient is employed to characterize the nature of ties in a networks \cite{newman2003mixing}. For example, in a large social network where individuals are characterized by their age, a positive assortativity would indicate that people of comparable age prefer to associate with each other, while a negative assortativity would indicate the opposite. The assortativity coefficient ($a$) is defined such that it lies between $-1$ and $1$. When $a=1$, the network is perfectly assortative, and when $a=-1$, the network is called completely disassortative (see section \ref{Appendices} for details).  In our case, we use the region of origin for the families to calculate $a$ for the network of families and the birth regions of the individuals for the case of the kin network. We obtain, $a=0.535\pm 0.023$ for the network of families and $a=0.277\pm 0.008$ for the kin network. These values indicate the role played by space in structuring the network and overall it reflects the fact that individuals are more prone to marry within %with 
the same region. However, the different regions have their own characteristics, which we investigate in the following fashion. We remove a particular region and calculate the assortativity coefficient ($a^{*}$) using the links for rest of the 17 regions. The difference, $a^*-a$ (shown in Fig. \ref{assortativity}) allows us to gauge the influence of that region on the overall assortativity of the network. A large positive value would indicate that the nodes from the region that is excluded in the process of calculation, in general, lowers the value of the coefficient when included in the calculation. It also means that the nodes themselves have a less tendency to associate with each other when compared with a typical node in the whole network. This phenomena appears to be strongest in cases of region 01 (Uusimaa) and 17 (Pohjois-Pohjanmaa). A contrasting observation is found for the case region 15 (Pohjanmaa) where we detected the presence Swedish speaking families. 

\begin{figure}
\centering
\includegraphics[width=0.95\columnwidth]{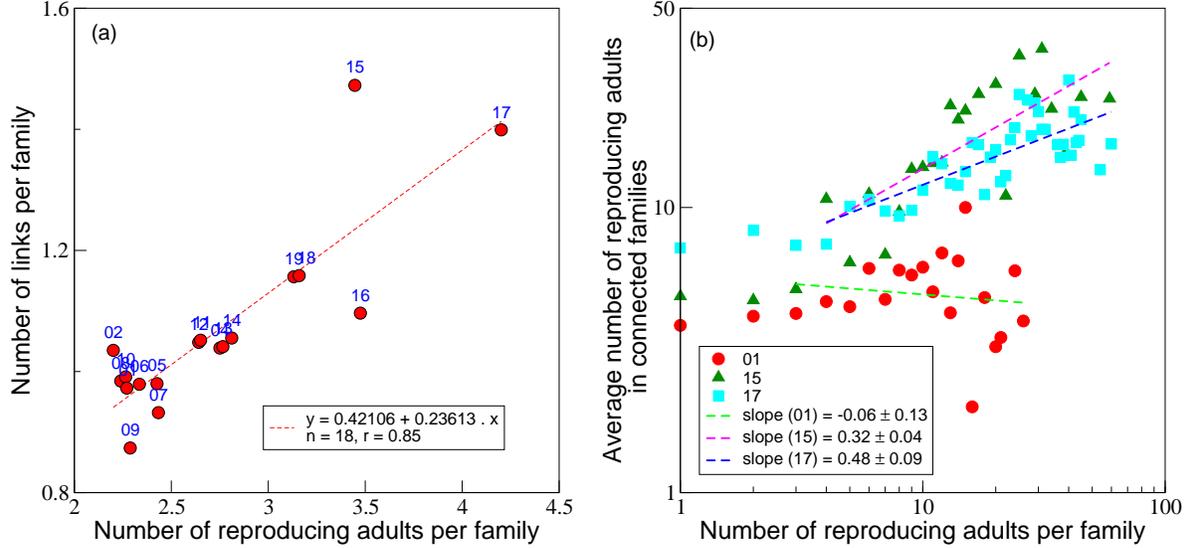}
\caption{{\bf Aspects of the network as functions of the number of reproducing adults in the family.}(a) The variation of number of links per family in the network of families with number of reproducing adults  per family. The regions to which each of the points correspond are denoted by the labels. The dashed line is a regression fit. (b) For a family with a given number ($x$) of reproducing adults  in a given region, we count the total number of reproducing adults in all the families (belonging to the same region) connected to it, and then divide by the number of connections (gives $y$). For each $x$  we calculate the mean $y$. These values are shown as scatter plots for three different regions, namely, Uusimaa (circles), Pohjanmaa (triangles) and Pohjois-Pohjanmaa (squares). The dashed lines are regression fits, with slopes provided in the legend. Unlike in the case of Uusimaa, for Pohjanmaa and Pohjois-Pohjanmaa the slopes are significantly greater than zero.}
\label{efs}
\end{figure}

\section{Summary and discussion}

In this study we have investigated the patterns of families in a contemporary European population through a network constructed from data on extended families in Finland. We consider the families as nodes and the links result from joint parenting of children by individuals from the families. We have characterized the structural properties of network of families and explored the transitivity and assortativity of the network with regards to the region of origin and linguistic identity. Using a large Finnish register data, we could link index persons with their parents, siblings, children, and grandchildren and further identify the links joining these extended families through marriage and reproduction. The results show that the sampled network is a collection of many disjoint components with the largest connected component including 8 per cent of all linked families and 14 per cent of all the detected parental unions.%links through marriage and reproduction joining these extended families. Results showed that the sampled network was a collection of many disjoint components with the largest connected component including 8 per cent of all linked families and 14 per cent of all parental unions detected. 

As could be expected, the capital Helsinki and its surroundings in Uusimaa region ($01$), which has the highest population size and density of all the regions in Finland, dominates with regards to the frequency of families. The total number of families (with or without links) that originate from this region make up clearly the largest fraction of the total set of families. However, in the case of the network composed of families linked with other families, the pattern is changed. Of families that have links, the largest proportion ( $13\%$) originate from the region of Pohjois-Pohjanmaa ($17$). This region and its regional neighbours are also known for their comparatively high fertility.  The families with linkages produce a set of connected components with the largest connected component (LCC) being made up of $957$ families and $1,211$ links. The number of individuals in the dataset (around $0.7$ million, counting the number of persons in each family) is of around $10\%$ of the Finnish population (just over $5$ million). To our knowledge this is the first time the presence of one single connected network between families in such a representative population sample has been documented.

We have also found that patterns of connectivity in the network are influenced by the regions, in which the families are rooted. This finding is in line with a number of %the many existing 
studies showing that the region of origin remains important for sociality of Europeans today \cite{kolk2014multigenerational}. Intensities of internal migration are known to be higher in Finland and Scandinavia compared to Southern European countries \cite{bell2015internal}, yet although a large proportion of Finns migrate to another region during the time of young adulthood, many eventually move back or closer to their region of origin once they have children themselves or after retirement \cite{tervo2000suomen}. Interestingly, the first and second largest connected components were predominantly populated by families rooted in a few specific regions.
Furthermore, the kin graphs corresponding to the cores of the four largest connected components were all dominated by one of the two national languages, Finnish or Swedish, the latter spoken by $6\%$ of the total population but represented much more in some regions. % overrepresented in some regions. 
The fact that cultural homophily, in terms of religion and language, plays a major role becomes evident in our investigation of the presence of transitive %by investigating for the presence of transitive 
relations between families. We found that the concentrated presence of a minority group of people with Swedish being their mother tongue, is reflected in the proliferation of triangles. Thus the majority of members in the families in the transitive core part of the (one of the two) second largest connected component came from the region 15 (Pohjanmaa) and were Swedish speaking. It is known that $40\%$ of the Swedish speaking population of Finland resides in this particular region. Furthermore, the kin cores of this particular connected component has the largest proportion of degree and clustering as well as a higher estimate of the assumed genetic relatedness than the largest connected component has. The patterns revealed through the structure of the network is consistent with the genetic clustering found in the Swedish speaking population of Pohjanmaa \cite{hannelius2008population}. 

In the network of families, the ties are constituted by two individuals of opposite sex who jointly parent one or more children (the first born is included in the kin network). In general, each family has a number of reproducing individuals and some become part of the linkages in the sampled network when the family of the opposite sex partner is also present in the data. Therefore, under a simplistic description, the larger the family (and hence larger is the number of reproducing adults), the more is the chance of this family to have a link.  The plot of the number of links per family against the number of reproducing adults for the different regions is shown in Fig. \ref{efs}(a). Here we have taken into account all families, even those that could not be linked. This approach is expected to reduce the sampling bias. The linear correlation is $r=0.85$ and a fit suggests a linear relationship. As discussed, the region 17 (Pohjois-Pohjanmaa) and its neighbours (15 (Pohjanmaa), 16 (Keski-Pohjanmaa), 18 (Kainuu) and 19 (Lappi)) on account of having a higher fertility are positioned towards the right on the horizontal axis. A critical view on contraception is likely to have contributed to this effect, which is a consequence of religious identities of families like being members of Lutheran sects such as the Laestadians \cite{terama2010regional} in this region. We cannot be more concrete at this point as our data does not include information regarding the religious affinities. In contrast to this, the region 01 (Uusimaa) and other regions in the south are found to have lower fertility. Although, region  17 (Pohjois-Pohjanmaa) appears to have the largest of the families, its deviation from the linear relationship is negligible, whereas region 15 (Pohjanmaa) and 16 (Keski-Pohjanmaa) show larger deviations. In the case of the region 16 (Keski-Pohjanmaa) we observed families to be mostly linked to families from neighbouring regions.

Correlations in the connectivity pattern for families in the region 17 (Pohjois-Pohjanmaa) can not be solely judged by the aspect of regional assortativity and large sized families resulting from higher fertility rate. There appears to be a tendency for the large families to get connect to each other. This is shown in Fig. \ref{efs}(b). For a family of a given size (measured in terms of the number of reproducing adults) we calculate the average size of the connected families. This is similar to the nearest neighbours average connectivity, which is used to quantify the degree correlations in networks \cite{pastor2001dynamical}. A positive slope corresponding to the region 17 (Pohjois-Pohjanmaa) indicates the presence of such correlations. Similar correlation is also present in the region 15 (Pohjanmaa). For the rest of the regions we did not find any significant correlation. The case of the region 01 (Uusimaa) is illustrated where the slope is not different from zero. This kind of ``degree assortativity'' originating likely from religious reasons in addition to the regional assortativity could be the reason for them dominating in the largest connected component  \cite{newman2002assortative}). In fact it was demonstrated in \cite{newman2002assortative} that when such assortativity is high a ``core group'' is formed by high degree nodes on which a largest connected component grows but contrary to expectations does not grow steadily and does not extend into the rest of the network. The scenario is very similar to our case, and such high assortativity and resulting impedance in the growth of the largest component could additionally imply that the true underlying network (from which the data is sampled) is not a small world \cite{gomez2004local,small2008scale}. It may be surprising to a certain extent as the fragments listed in Table \ref{tab:kin-stat} are small worlds.

In sum, the general patterns of linkages found within this representative sample of a national population are indicative of a high assortativity in the network of families. Both the region of birth and the language appear to function as cultural attractors in the network and increase the clustering and transitivity. We can distinguish between two patterns of regional effects in this network, either showing ``metropolitan'' family linkages or the ``cultural'' family linkages. %We distinguish them as ``metropolitan'' family linkages and the cultural family linkages. 
The metropolitan families are to be found in region around the capital and they %. They 
are mostly part of smaller clusters and many of these families could not be linked to other families in this population sample. These families are present in large numbers and appear to be overwhelmingly linked to families originating from the other regions. Migration of population to the more industrialized southern regions of Finland results into the lowering of assortativity and network transitivity.
The cultural linkages are found among families from the Pohjois-Pohjanmaa region (17) as well as other western and northern regions of Finland. Here, the regional and linguistic identity seems to result in a strong regional connectivity in terms of family ties.

\bibliography{manuscript.bbl}

\section*{ACKNOWLEDGEMENTS}
KB, DM, AG, and KK acknowledges financial support by the Academy of Finland Research project  (COSDYN)  No.   276439  and  EU  HORIZON  2020  FET  Open  RIA  project  (IBSEN)  No.
662725.  J.K. was partially supported by H2020 FETPROACT-GSS CIMPLEX Grant No.  641191.

\section*{Appendices}
\label{Appendices}

\subsection*{Assortativity coefficient}
Following \cite{newman2003mixing}, we construct a symmetric matrix $\left\lbrace e_{ij} \right\rbrace$, where $e_{ij}$ is the fraction of edges connecting nodes belonging to region $i$ and region $j$. The edges connecting nodes of the same type are counted twice. The matrix satisfies, $\sum_{ij}e_{ij}=1$, and $\sum_{j}e_{j}=c_i$, where $c_i$ is the fraction of type of end of an edge that is attached nodes belonging to region $i$. The assortativity coefficient is given by:
\begin{equation*}
a=\frac{\sum_ie_{ii}-\sum_ic_i^2}{1-\sum_ic_i^2},
\end{equation*}
with the error estimate ($\sigma_a$) being,
\begin{equation*}
\sigma_a^2=\frac{1}{M}\frac{\sum_ic_i^2+\left[\sum_ic_i^2\right]^2-2\sum_ic_i^3}{1-\sum_ic_i^2},
\end{equation*}
where, $M$ is total number of edges in the network.
\label{appendices}
\renewcommand{\thefigure}{A\arabic{figure}}
\setcounter{figure}{0}

\begin{figure}
\centering
\includegraphics[width=0.85\columnwidth]{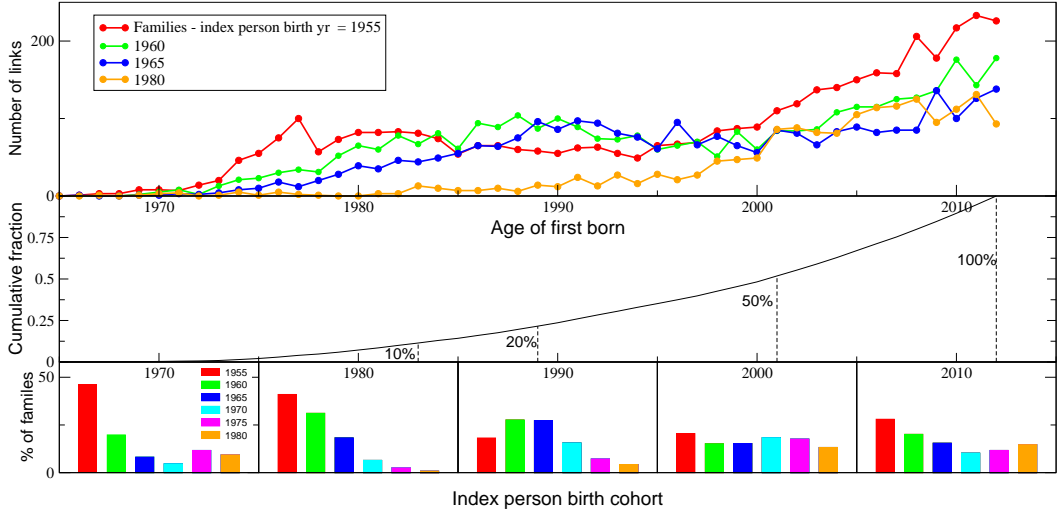}
\caption{{\bf Growth of the network in time.} (Top) The number of links that appear on a given year (birth year of the first born from a parental union) having one family (or both) with a year of reference indicated in the legend (four out of the six cases are shown). (Middle) The cumulative fraction of links as they appear in time. (Bottom) Relative presence of families from the six years of reference in five non-overlapping  periods -- $1966$-$1975$, 1976-1985, 1986-1995, 1996-2005, and 2006-2012. The variation in the height of the bars reveal the periodicity that is present in the contribution.}
\label{link-year}
\end{figure}

\begin{figure} 
\centering
\includegraphics[width=0.75\columnwidth]{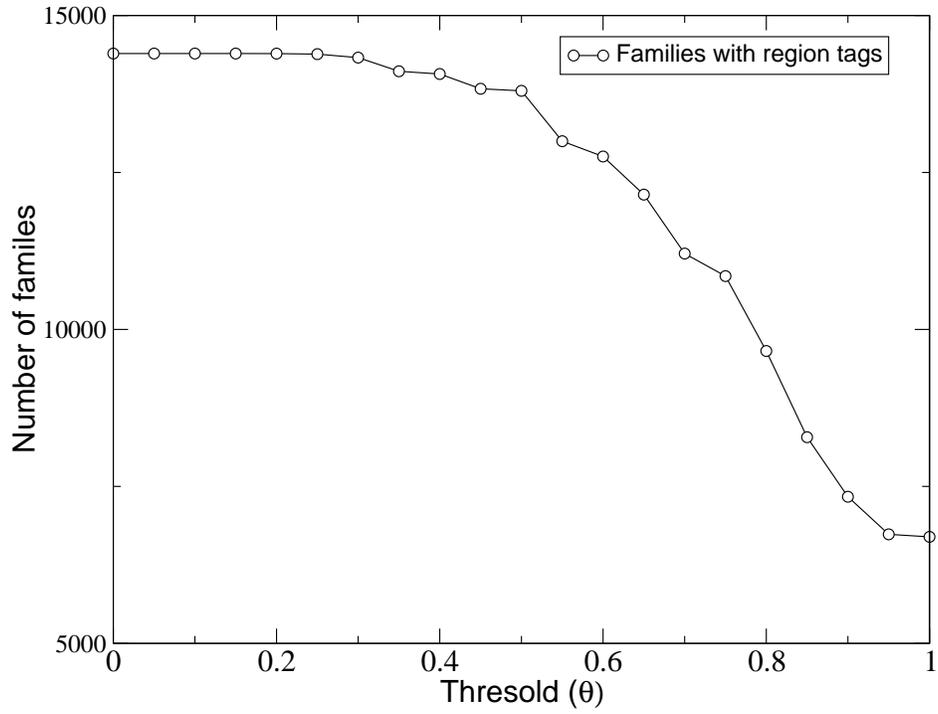}
\caption{Number families that could be included in the study based on the threshold $\theta$. A given family is assigned a region of origin $m$ if at least a fraction $\theta$ of the reproducing adults have the region of birth as $m$.}
\label{fig-threshold-theta}
\end{figure}

\begin{figure} 
\centering
\includegraphics[width=0.75\columnwidth]{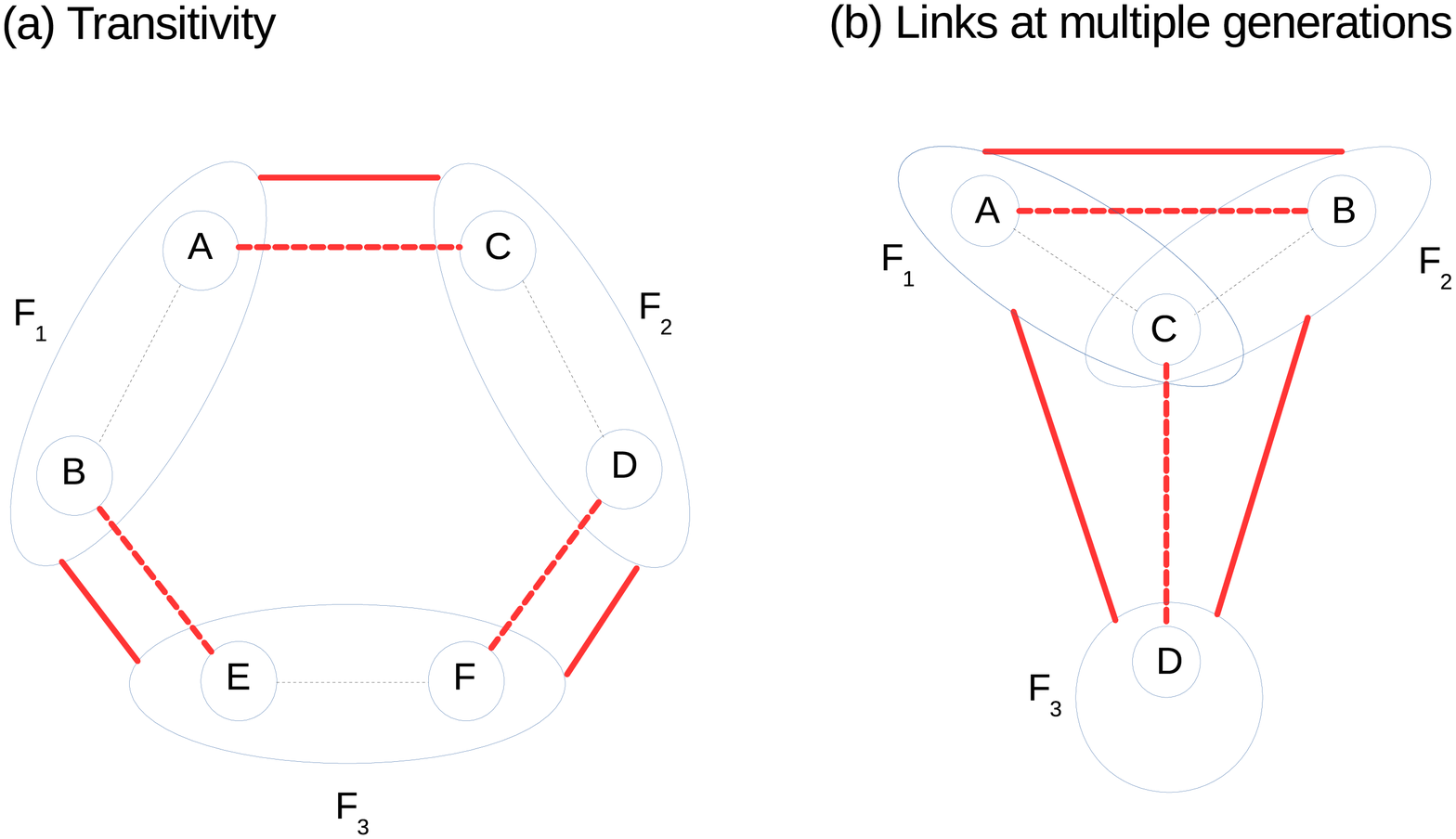}
\caption{{\bf Possible types of triangles present in the network of families.} In the above diagram, F$_1$, F$_2$ and F$_3$ denote families, and A, B, C, D, E and F denote individuals. Two individuals are joined by a thin dashed line when they belong to the same family. A thick dashed line between a pair of individuals belonging to different families indicate male-female partners in parental unions. The thick solid lines indicate connections between families as a result of the parental unions. In cases of both (a) and (b), F$_1$, F$_2$ and F$_3$ form closed triplets or triangles, although  nature of the triangulations are different. (a) Transitivity: A possible example is the following. The individuals A and B are brothers. A marries C, while B marries E. C and D are sisters, while E (female) and F (male) are cousins. F and D get married. The triangle formed by the families (F$_1$,F$_2$,F$_3$) is transitive because each link in the network would uniquely correspond to a marital relationship (and also result in one or more offspring, for consideration in our study). (b) Links at multiple generations: This kind of triangulation occurs when an offspring becomes a parent. For example, A (male) and B (female) get married and C is born. By definition, C belongs to both the F$_1$ (paternal family) and F$_2$  (maternal family). C gets married to D from family F$_3$. In this case, both links (F$_1$,F$_3$) and (F$_2$,F$_3$) result from the link between C and D.}  
\label{triads}
\end{figure}

\begin{figure} 
\centering
\includegraphics[width=0.9\columnwidth]{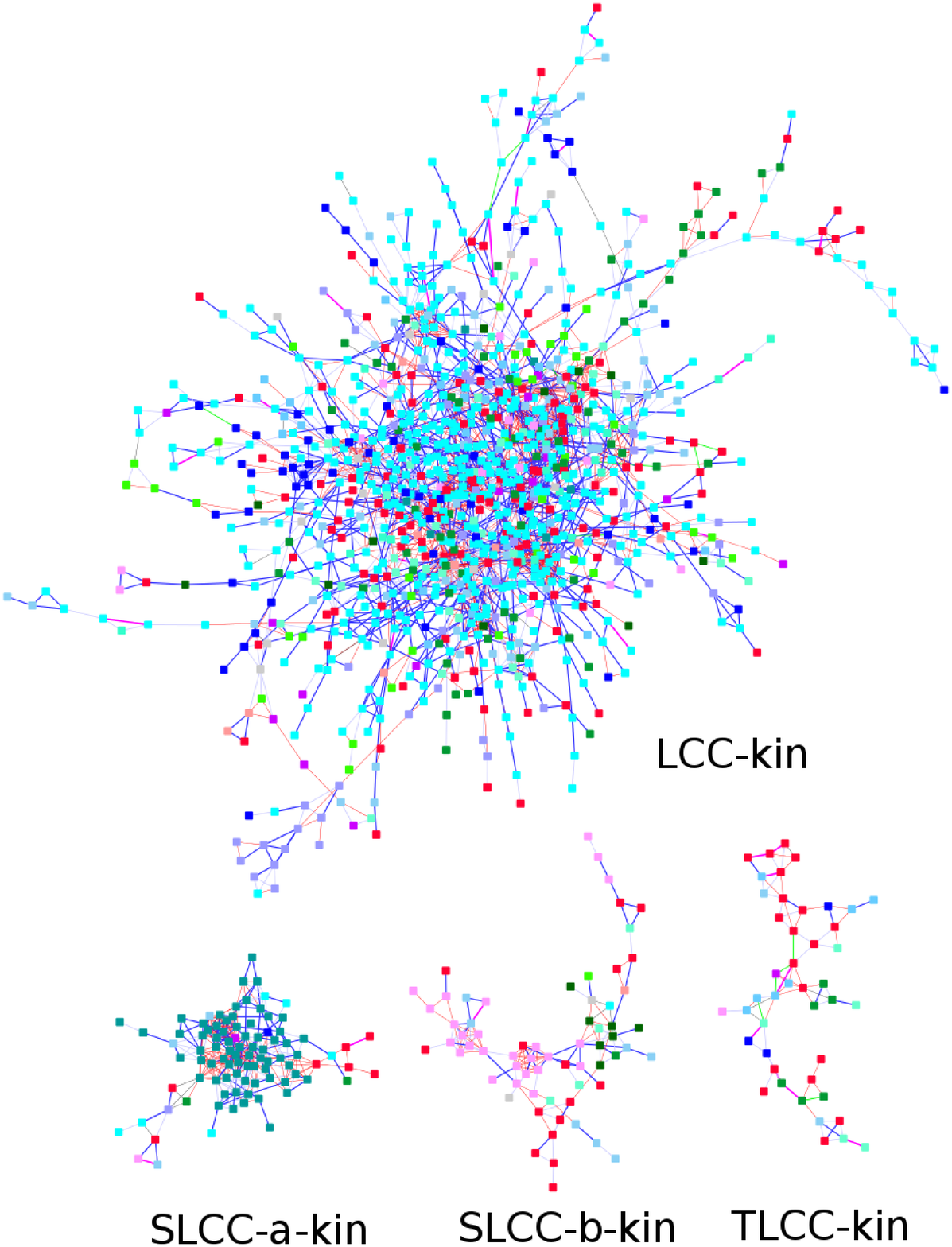}
\caption{{\bf The kin graphs derived from the four largest components of the network of families.} The colours represent the birth region of the individuals and scheme is the same as that in Fig. \ref{network-paper}.}
\label{kin-network}
\end{figure}

\end{document}